# NMRPy: a novel NMR scripting system to implement artificial intelligence and advanced applications

Zao Liu[1,2], Kan Song[2*], Zhiwei Chen[3*]

*Corresponding authors:

Zhiwei Chen

Email: chenzhiwei@xmu.edu.cn

Institute: Department of Electronic Science, Fujian Provincial Key Laboratory of Plasma and Magnetic Resonance Research, Xiamen University, Xiamen 361005, P. R. China

Kan Song

Email: songkan@qone-inst.com

Institute: Zhongke-Niujin MR Tech Co. Ltd, Wuhan 430075, P. R. China

Full list of author information is available at the end of the article

**Abstract**

**Background:** Software is an important windows to offer a variety of complex instrument control and data processing for nuclear magnetic resonance (NMR) spectrometer. NMR software should allow researchers to flexibly implement various functionality according to the requirement of applications. Scripting system can offer an open environment for NMR users to write custom programs with basic libraries. Emerging technologies, especially multivariate statistical analysis and artificial intelligence, have been successfully applied to NMR applications such as metabolomics and biomacromolecules. Scripting system should support more complex NMR libraries, which will enable the emerging technologies to be easily implemented in the scripting environment.

**Result:** Here, a novel NMR scripting system named "NMRPy" is introduced. In the scripting system, both Java based NMR methods and original CPython based libraries are supported. A module was built as a bridge to integrate the runtime environment of Java and CPython. It works as an extension in CPython environment, as well as interacts with Java part by Java Native Interface. Leveraging the bridge, Java based instrument control and data processing methods can be called as a CPython style. Compared with traditional scripting system, NMRPy is easier for NMR researchers to develop complex functionality with fast numerical computation, multivariate statistical analysis, deep learning etc. Non-uniform sampling and protein structure prediction methods based on deep learning can be conveniently integrated into NMRPy.

**Conclusion:** NMRPy offers a user-friendly environment to implement custom functionality leveraging its powerful basic NMR and rich CPython libraries. NMR applications with emerging technologies can be easily integrated. The scripting system is free of charge and can be downloaded by visiting http://www.spinstudioj.net/nmrpy.

**Background**

Since it was discovered in the 1940s, nuclear magnetic resonance (NMR) has achieved important applications in many fields. Emerging domains such as metabolomics[1] and biomacromolecules[2, 3] require automated, flexible data acquisition and processing for NMR software. With the rapid development of methodology field, new technology has played an important role in modern NMR. For example, artificial intelligence[4-6] is a researching hotspot to solve complex chemical and biological problems. Non-uniform sampling[7] by deep learning has greatly accelerated multidimensional experiments of biological samples; Chemical shift[8] and protein structure[9] prediction allow users to predict NMR spectrum and 3D shape of a protein without long time experiment; Multivariate analysis[10] for metabolomics uses multivariate statistical algorithms to reveal the relationship between the quantity of metabolites and biological or medical problems. Novel NMR methods and processing algorithms have to be conveniently implemented and integrated into the software. Scripting system offers a powerful and open environment that allows users to write scripting programs to implement their own instrument control and data processing methods. Therefore, the issues of providing more powerful utility libraries and flexible programming features are significant for the design of NMR scripting system.

To enhance the performance of the scripting system, most popular scripting languages have been applied in NMR software. In general, existing scripting system can be divided into two types. For the first type, scripting system runs as an extension of the main program which is compiled with another computer language. Most commercial NMR software belongs to the first type. MAGICAL(MAGnetics Instrument Control and Analysis Language)[11] applied in software VnmrJ is developed based on "shell" scripting language which is native in UNIX like operating systems, the macros of MAGICAL support complex pulse sequence and custom commands; Jython[12], Tcl[13] and AU program[14] are supported in TopSpin[15] to undertake different tasks. Jython and Tcl are standard scripting languages. AU program is based on C language and macros, and it has to be compiled in GNU environment; Mnova[16] uses native scripting language of Qt[17] library named "QtScript" to call powerful NMR algorithms of C++ based programs; ACD/Spectrus Processor[18] supports several standard scripting language(such as BasicScript, PascalScript, JavaScript and C++Script) to perform data processing and analysis sequentially. For the second type, the whole NMR software is built by a scripting language. MatNMR[19], jsNMR[20], rNMR[21] and nmrglue[22] use MATLAB, JavaScript, R language[23] and CPython[24] as scripting languages respectively for NMR data post-processing, taking advantage of their powerful scientific computing libraries and chart display. However, there are some essential limitations for both types. Scripting system in the first type is limited in advanced algorithm such as fast numerical computation and artificial intelligence; For the second type, there are some critical disadvantages in execution efficiency and implementing complex graphical user interface.

To overcome above challenges and propose a more powerful and flexible scripting system, it is better to combine scripting and compiling computer languages to take advantage of both languages. To implement modern requirements of NMR methods, complex logical control and advanced computation especially artificial intelligence libraries have to be supported in scripting system. As a free and open source scripting language, Python, not only has flexible syntax features, but also has formed a powerful language ecosystem due to many kinds of libraries. It has a variety of implementations with different programming language: Jython, IronPython[25], PyPy[26] and CPython. Jython, IronPython and CPython are respectively implemented with Java, C# and C programming languages, PyPy is implemented by JIT (Just-In-Time Compiler) to make it run faster. Among these Python implementations, only CPython has developed rich libraries and powerful ecosystem. It is an ideal choice as NMR scripting language; Java language[27, 28] is widely used in many fields due to its cross platform feature and powerful functionality, it is suitable to perform instrument control, data processing and display.

In this paper, a new NMR scripting system named "NMRPy" is introduced. By integrating CPython and Java language, it has the capabilities of offering rich CPython based libraries including scientific computation and artificial intelligence, as well as Java based conventional NMR functionality which has advantages in graphical user interface and interaction with users.

## Implementation

### Architecture

The scripting system(named "NMRPy") offers a flexible scripting environment to implement custom functionality for NMR users. Conventional instrument control and data processing are implemented by Java, and CPython has advantages in advanced numerical computation and artificial intelligence(e.g. NumPy[29], TensorFlow[30]). The significant issue of the scripting system is to build a bridge to connect Java and CPython, so that both Java-based NMR methods and third-party libraries of CPython are supported. CPython is developed by C language. It provides C extension to wrap C libraries for CPython as customized modules; In addition, Java virtual machine provides a mechanism named "Java Native Interface (JNI)"[31] to interact with C language. Through the interface, Java can call functions defined by C, and C also can visit all kinds of resources(e.g., classes, functions, objects) in Java environment. So, C language is an ideal bridge between Java and Python.

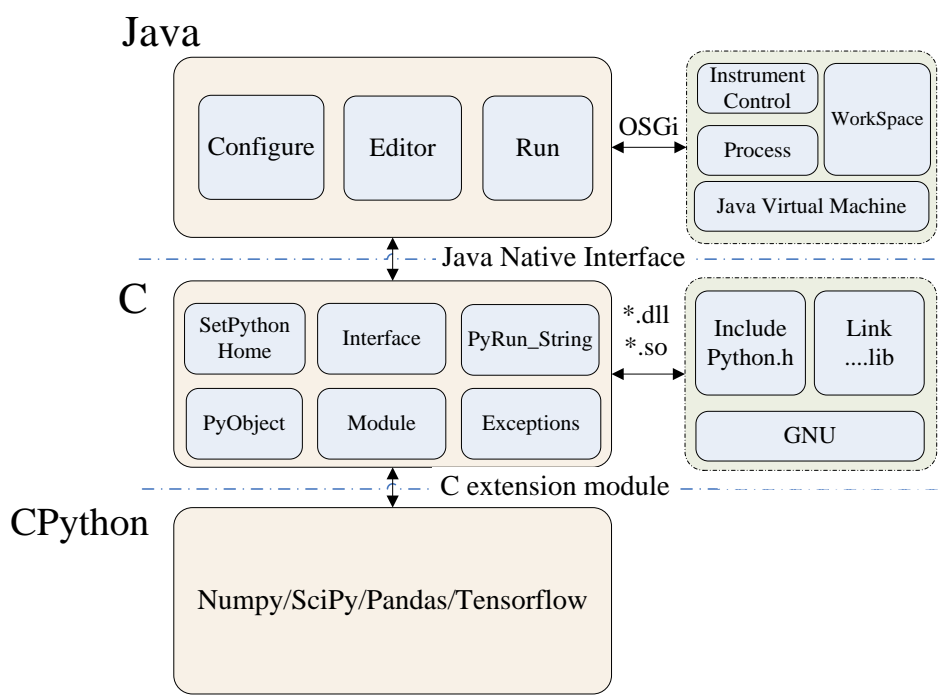

Figure 1. Architecture of the scripting system

The overall architecture of NMRPy is illustrated in Figure 1. According to the computer language, the entire scripting system consists of three components: Java part, C part and CPython part.

Java part is generally responsible for graphical user interface and providing interfaces for CPython, including basic configuration, scripting editor, interfaces of instrument control and data processing. Configuration can set the location of CPython libraries; Scripting editor offers script editing window, execution output, menus and toolbar; Interfaces for instrument control and data processing are implemented by OSGi (Open Service Gateway Initiative) which separates abstract interface from concrete business logic. Instrument control includes sample control, temperature control, tuning, locking, shimming, and data acquisition, etc. Data processing includes Fourier transform, phase correction, baseline correction, peak picking, integration. The NMR data of custom format is consisted by parameters, FID, spectrum, peak list, etc, it can be read by the Java interface provided for scripts. After some algorithm of CPython scripts, the processed data will be written back and saved to the disk. The scripts can be set to both blocking and non-blocking modes to ensure that the statements are executed in expected sequence.

C part is the bridge between Java and CPython part. JNI is the interaction interface defined by Java virtual machine (JVM). Through the interface, Java can call functions of C based libraries. If necessary, C part also can create Java objects and call Java functions. Based on C extension mechanism of CPython, C part can define customized modules for CPython, including initialization, exit, and methods exposed to CPython.

CPython part can define customized initialization and import methods for packages and modules, as well as offer various native libraries(e.g., NumPy, SciPy, TensorFlow). Initialization and import are the significant steps which make CPython interact with Java. For native libraries, NumPy and SciPy are usually used for fast numerical operations and scientific calculations. TensorFlow is widely used for deep learning. Non-uniform sampling and chemical shift prediction methods developed by deep learning can be easily integrated into the scripting system.

**Workflow**

The workflow of NMRPy can be divided into two stages: initialization and execution.

Initialization is a significant step to build the scripting environment. First, scripting system will configure the path of CPython libraries. Second, CPython will install an importer hook and insert it to "sys.meta_path". The importer hook will define methods "find_module" and "load_module" to tell CPython how to find and load Java packages.

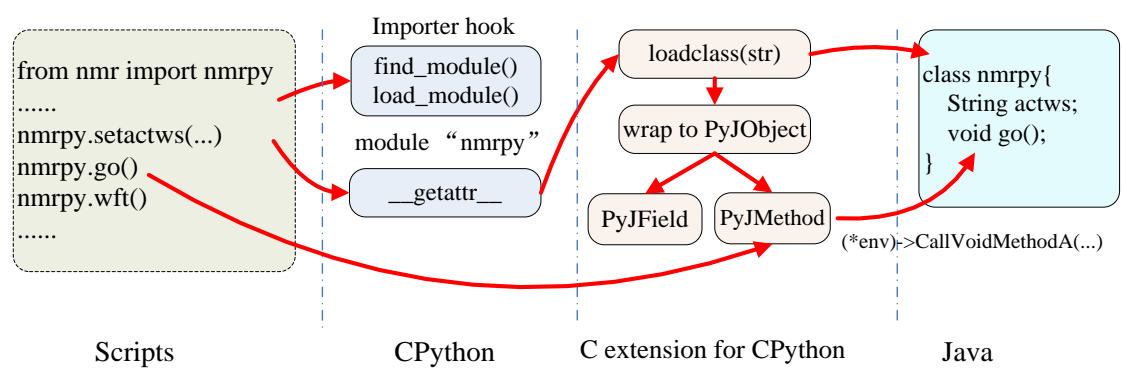

Figure 2. Execution workflow of the scripting system

The workflow of execution is illustrated in Figure 2. Import and interpretation are the significant issues in the stage of execution. When "import" statement is called, scripting system will search the expected package from "sys.modules". If the package is found, it indicates that the package has been loaded by CPython; If not found, CPython environment will find the importer hook to invoke "find_module" and "load_module" methods to load module "nmrpy" which is used to interact with Java resources. In addition, module "nmrpy" will be added to "sys.modules". Module "nmrpy" has implemented method "__getattr__" to define submodules of "nmrpy" as the attributes for packages and classes in Java environment. The methods (wraped to "PyJMethod") and fields (wraped to "PyJField") of Java objects will be wraped as the attributes of an object of CPython. CPython allows "PyJMethod" to implement custom method execution by defining attribute "tp_call" of "PyTypeObject", and allows "PyJField" to implement custom getting and setting style by defining attributes "tp_getattro" and "tp_setattro" of "PyTypeObject". When NMR command "go" is executed in scripts, "PyJMethod" will invoke corresponding Java method by the Java Native Interface for methods. Therefore, CPython interpreter can recognize Java objects as conventional native CPython objects, as well as calling Java methods freely.

Exception and memory management are other significant issues for scripting system. JNI allows C part to throw C based exceptions to Java part, then Java part will catch exceptions and back traces. For memory management, Java part can reclaim the runtime useless memory automatically due to garbage collection of JVM, so there is no need to release memory manually in conventional condition; C part has to release memory manually, both JNI and C extension part offer corresponding methods to release useless memory to avoid memory leaks.

**Results**

NMRPy provides an user friendly graphical user interface. The screenshot of NMR scripting editor is illustrated in figure 3. The menus and tools offer conventional functionality for file access and text editing. The editor gives a variety of example scripts such as automated experiment, singular value decomposition (SVD), data plotting. For script codes, key words of CPython can be marked as a highlighted style. Code comments and strings can be displayed as special colors. The bottom region can show the outputs during the execution of scripts. The reported errors, warnings, and exceptions can prompt users to deal with problems during the execution of script programs.

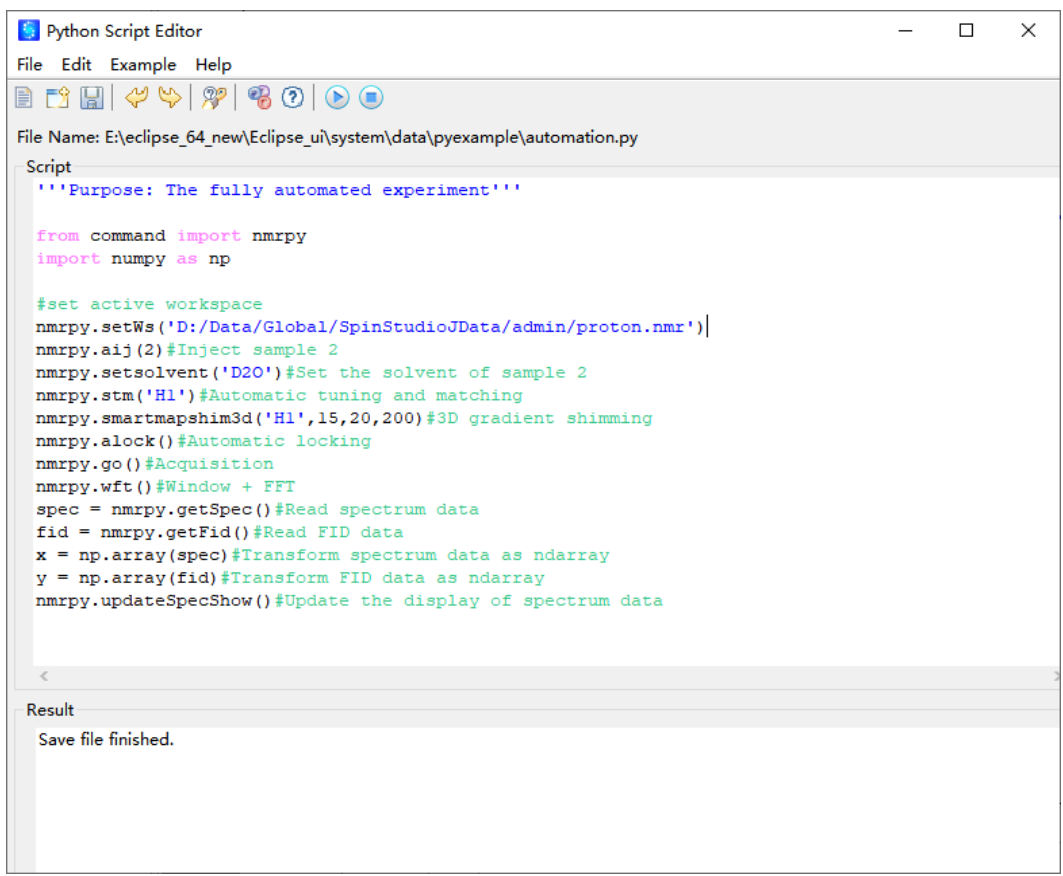

Figure 3. The screenshot of NMR scripting editor

The scripts offers general functionality such as instrument control, data processing as well as native CPython libraries. Typical scripting functions are described as Table 1.

Table 1. General functions in NMRPy

| Function name | Description | Example |
| --- | --- | --- |
| Instrument control | | |
| aij(n) | Inject sample, sample number is "n". | aij(2) |
| alock() | Lock the field automatically. | alock() |
| stm(nucleus) | Automatic tuning and matching | stm('H1') |
| smartmapshim() | Make gradient shim map and do shimming. | smartmapshim() |
| smartshim() | Gradient shimming using exist field map. | smartshim() |

| searchshim(algorithm,evaluation,channels,iteration) | Searching better shim values with some algorithm. | searchshim('simplex', 'FIDArea', 'z1-z2',50) |
|---|---|---|
| vartemp(target,timeout) | Vary temperature to "target" celsius degree within "timeout" seconds | vartemp(35.5,240) |
| spin(target,timeout) | Rotate the sample with spin rate of "target" Hz within "timeout" seconds | spin(20,200) |
| setshimvalue(channel,value) | Set the shim value of shim coil in "channel" | setshimvalue('z1',1000) |
| go() | Start data acquisition | go() |
| Data processing | | |
| setactws(path) | Set active workspace | setactws('D:/1.nmr') |
| setparam(name,value) | Set the value of parameter of active workspace | setparam('ns',4) |
| getfid(path) | Get FID data of workspace whose storage path is "path" | getfid('D:/1.nmr') |
| getspec(path) | Get spectrum data of workspace whose storage path is "path" | getspec('D:/1.nmr') |
| setspec(path) | Get spectrum data of workspace whose storage path is "path" | setspec('D:/1.nmr') |
| wft() | Data processing with weighting and Fourier transform | wft() |
| Original CPython libraries | | |
| np.multiply(a,b) | Matrix multiplication | np.multiply(a,b) |
| np.median(a) | Median of array | np.median(a) |
| plt.plot(x,y) | Draw a curve | plt.plot(x,y) |
| scipy.optimize.curve_fit(func,x,y) | Curve Fitting | scipy.optimize.curve_fit(func,x,y) |

Instrument control is used to control NMR spectrometer's physical components such as auto sample changer, pneumatic, shimming and lock units. For data acquisition, scripts are allowed to set parameters of workspace, and then start the command of acquisition. Both blocking and non-blocking mode are supported. In blocking mode, the script will wait for the completion of invoked method until the maximum time is exhausted. In non-blocking mode, the script will invoke the method immediately and never wait for the completion of execution.

For data processing, scripts can call conventional processing methods such as linear prediction, Fourier transform, phase correction, baseline correction etc. Scripts can access free induce decay (FID) or spectrum from workspace, as well as update the spectrum display after some transformation or analysis.

In order to achieve powerful performance, most native libraries of CPython can be used in the scripting system, including NumPy, SciPy, matplotlib and TensorFlow, etc. FID or spectrum can be read from workspace as format "ndarray" or "Matrix" defined by NumPy, which supports user-friendly and efficient numerical manipulation; SciPy is a scientific computation library which can be used in parameter optimization and data denoising; Matplotlib library is a visualization library for FID or spectrum plotting; TensorFlow can be used to implement deep learning which is an emerging field in NMR.

To illustrate the performance of the scripting system, there are three following nuclear magnetic resonance methods as examples.

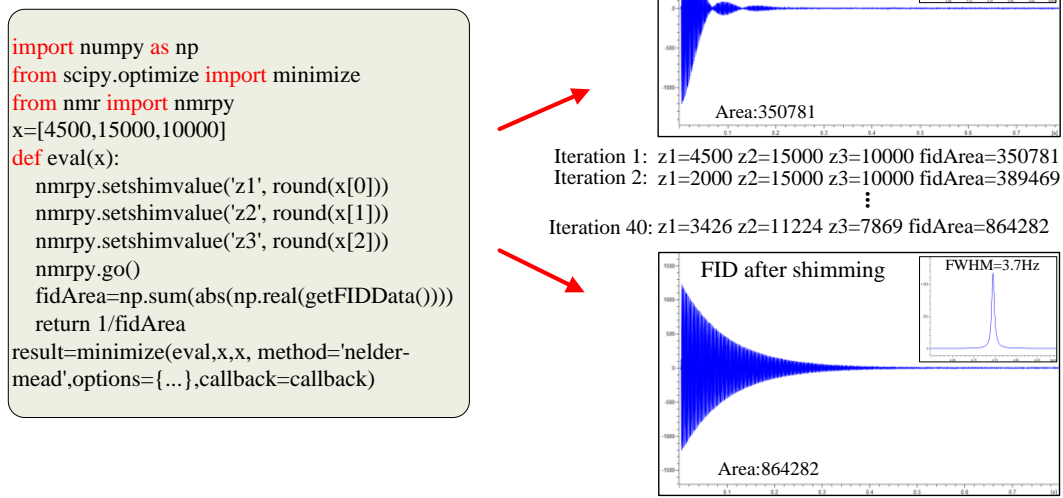

Figure 4. Searching shimming by FID area and simplex algorithm

The first script example for automatic searching shimming is illustrated in Figure 4. Automatic searching shimming aims to find the best shim values using multivariate optimization algorithm to get better field homogeneity, which is significant to improve the resolution of biological macro molecules. It needs iteratively real-time data acquisition and optimization analysis. NMRPy offers Java based methods for setting shim values and data acquisition. SciPy library provides optimization algorithm such as "Simplex"[32] to generate new shim values with an optimized searching path. As illustrated in figure 4, the experiment of searching shimming was performed on Zhongke-Niujin built 500 MHz QOne$^{Plus}$ NMR spectrometer. The test sample is 0.1mg/ml GdCl3 in D2O with 1% H2O. The evaluation criterion of field homogeneity is the area of FID, and the optimization algorithm is "Simplex". After 40 iterations, the area of FID is optimized from 350781 to 864282, and the half peak width of $H_2O$ peak is optimized from 15.8Hz to 3.7Hz.

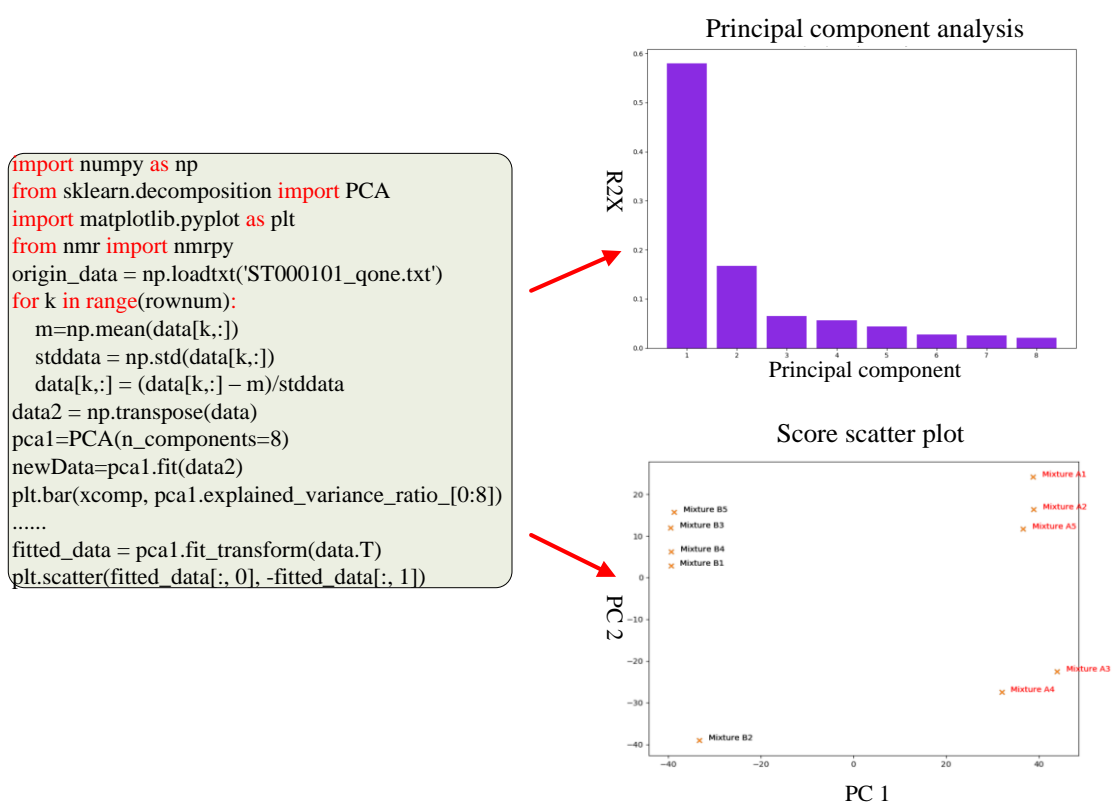

Figure 5. Principal component anlysis

The second example is for principal component analysis (PCA)[33] of metabolomic data. PCA is the significant step to find the difference of the bulk data by projecting the data onto multiple orthogonal components. As illustrated in figure 5, the data is an example from web site "metabolomics workbench", and the corresponding study ID is "ST000101"[34]. After Java based automatic phase, baseline correction and integration of binned spectral of 10 samples which differed by two groups of mixtures, the data set is analyzed by PCA which is implemented by the scripting system. The script of PCA includes reading the original data, calculating the average and standard deviation, principal component analysis, columnar and scattered data display. CPython libraries of NumPy, scikit-learn[35] and matplotlib are helpful to implement above requirements. The histogram gives the result of the 1st~8th principal components and their percentages. Among all the components, the first principal component can explain 58.02% information of the total samples. The score scatter plot shows the score value of each sample on the first and second principal component. Obviously, on the first dimension of principal component, the sample is divided into two groups A and B, which is consistent with the metabolite composition of the samples. The data of score scatter plot is also the same with the published result of the study.[36]

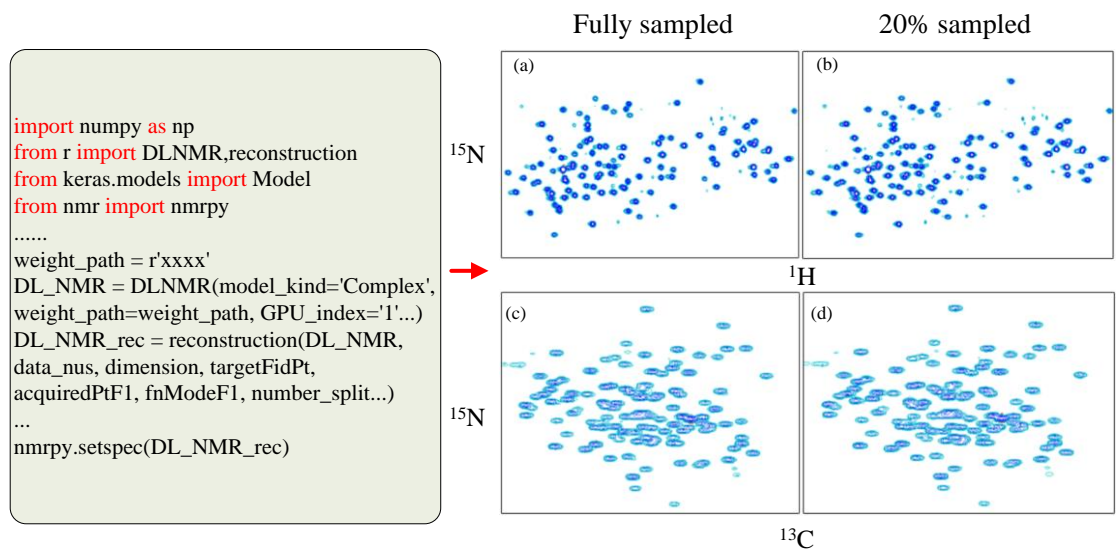

Figure 6. Non-uniform sampling by deep learning

The last example is about deep learning NMR(DLNMR)[7] for Non-uniform sampling (NUS)[6, 37, 38]. The scripts will reconstruct the spectra with a smart dense convolution neural network (DCNN)[7], which have been trained with simulated or acquired data. DCNN requires a variety of libraries to implement convolution and data fitting. "NMRPy" can integrate all the related libraries (such as TensorFlow, keras, cuDNN) in CPython environment. As is illustrated in Figure 6, a 3D HNCO NMR data of Azurin (molecular weight is 14kDa) with full sampling was downloaded from the MddNMR website http://mddnmr.spektrino.com. (a) and (c) are the sub-regions of projections on planes of $^{15}$N-$^{1}$H and $^{15}$N-$^{13}$C for the fully sampled 3D spectrum , which is reconstructed by fast Fourier Transform; (b) and (d) are the corresponding result of reconstruction result of only 20% sampling rate, which is reconstructed by DLNMR. The reconstructed spectrum by DLNMR is almost the same as the fully sampled one. For the 3D NMR, the achieved acceleration factor of 5 in NUS implies that the experimental time can be reduced from 22.4 hours to 4.48 hours. In addition, the heating effect of long time RF exciting also can be significantly reduced.

The above examples involve the advanced functionality: Simplex searching, principal component analysis, and dense convolution neural network. Comparing with traditional scripting system, NMRPy is easier for NMR researchers to develop complex NMR methods and applications.

## Conclusion

NMRPy is a novel NMR scripting system which can offer general functionalities for instrument control, data processing and original CPython libraries. Conventional instrument control and data processing are implemented by Java programming language, original CPython libraries are helpful for advanced algorithm such as fast numerical computation, artificial intelligence etc. More advanced NMR functionality such as chemical shift and protein structure prediction will be integrated in the future.

NMRPy can be downloaded free of charge by visiting the website: http://www.spinstudioj.net/nmrpy.

## Abbreviations

NMR: Nuclear Magnetic Resonance; NUS: Non-uniform sampling; MAGICAL: MAGnetics Instrument Control and Analysis Language; JIT: Just-In-Time Compiler; FID: Free Induce Decay; JNI: Java Native Interface; OSGi: Open Service Gateway Initiative; DCNN: Dense convolution neural network; PCA: Principal Component Analysis; SVD: Singular Value Decomposition; DLNMR: Deep learning NMR.

Declarations:

## Acknowledgement

We gratefully acknowledge Mr. Rui Chen, Mr. Rui Cao, Mr. Xuedong Zheng and Ms. Qingyuan Li for their patient programming assistance. We are very appreciated for Prof. XianZhong Yan's fruitful discussion.

**Authors' contribution**

Zao Liu finished the programming work and wrote the manuscript. Kan Song designed the examples of all scripts. Zhiwei Chen designed the main framework of scripting system. All authors have read and approved the final manuscript.

**Funding**

No funding was obtained for this study.

**Availability of data and materials**

All the programs, script examples can be downloaded by visiting web site: http://www.spinstudioj.net/nmrpy. The original data for PCA analysis is from https://www.metabolomicsworkbench.org/data/pca/show_metabolite_pca_NMR.php. The NMR example data of sample Azure for NUS is downloaded from http://mddnmr.spektrino.com.

**Ethics approval and consent participate**

Not applicable.

**Consent for publication**

Not applicable.

**Competing interests**

Not applicable.

**Author details**

[1]State Key Laboratory of Magnetic Resonance and Atomic and Molecular Physics, Wuhan Center for Magnetic Resonance, Wuhan Institute of Physics and Mathematics, Innovation Academy for Precision Measurement Science and Technology, Chinese

Academy of Sciences, Wuhan, 430071, P. R. China

[2]Zhongke-Niujin MR Tech Co. Ltd, Wuhan 430075, P. R. China

[3]Department of Electronic Science, Fujian Provincial Key Laboratory of Plasma and Magnetic Resonance Research, Xiamen University, Xiamen 361005, P. R. China